\documentclass[]{elsarticle}
\newpageafter{abstract}

\usepackage[utf8]{inputenc}
\usepackage{amsmath,amsfonts,amssymb,amsthm,bm}
\usepackage{booktabs}
\usepackage{graphicx}
\usepackage{tikz}
\usetikzlibrary{external}
\usepackage{pgfplots} 
\usepackage{pgfgantt}
\usepackage{pdflscape}
\pgfplotsset{compat=newest} 
\pgfplotsset{plot coordinates/math parser=false}
\usepackage{array}
\usepackage[obeyspaces,hyphens]{url}
\usepackage{adjustbox}
\usepackage{subcaption}
\newlength\fwidth
\newlength\fheight
\usepackage{acronym}
\acrodef{OFO}{Online Feedback Optimization}
\acrodef{MPC}{Model Predictive Control}
\acrodef{OPF}{Optimal Power Flow}
\acrodef{MIQP}{mixed integer quadratic optimization problem}
\newcommand{\bbR}{\mathbb{R}}
\DeclareMathOperator{\diag}{diag}

\usepackage[hyperindex=true, %
  pdftitle={Subtransmission Grid Control via Online Feedback Optimization},%
  pdfauthor={Lukas Ortmann, Jean Maeght, Patrick Panciatici, Florian D\"{o}rfler, Saverio Bolognani},%
  colorlinks=true,%
  pagebackref=false,%
  plainpages=false,%
  pdfpagelabels,%
  linkcolor=black,%
  citecolor=black,%
  filecolor=black,%
  urlcolor=black%
  ]{hyperref}
  \usepackage{xcolor}

\begin{document}

\begin{frontmatter}

\title{Subtransmission Grid Control\\via Online Feedback Optimization\tnoteref{t1}}

\tnotetext[t1]{The research leading to this research has been funded by the Swiss Federal Office of Energy through the project “UNICORN” (SI/501708).}

\author[1]{Lukas Ortmann\corref{cor1}}
\ead{lukas.ortmann@ost.ch}
\cortext[cor1]{Corresponding author}

\author[2]{Jean Maeght}
\ead{jean.maeght@rte-france.com}

\author[2]{Patrick Panciatici}
\ead{patrick.panciatici@rte-france.com}

\author[3]{Florian D\"{o}rfler}
\ead{dorfler@ethz.ch}

\author[3]{Saverio Bolognani}
\ead{bsaverio@ethz.ch}

\affiliation[1]{
    organization={OST – Eastern Switzerland University of Applied Sciences},
    city={Rapperswil},
    country={Switzerland}
}

\affiliation[2]{
    organization={R{\'e}seau de Transport d'{\'E}lectricit{\'e} (RTE)},
    city={Paris},
    country={France}
}

\affiliation[3]{
    organization={Automatic Control Laboratory, ETH Zurich},
    city={Zurich},
    country={Switzerland}
}

	\begin{abstract}
    The increasing electric power consumption and the shift towards renewable energy resources demand for new ways to operate transmission and subtransmission grids. Online Feedback Optimization (OFO) is a feedback real-time control method that can be employed to enable optimal operation of these grids. Such controllers can maximize grid efficiency (e.g., minimizing curtailment) while satisfying grid constraints like voltage and current limits. The OFO control method is tailored and extended to handle discrete inputs and it is explained how to design an OFO controller for the subtransmission grid. A novel benchmark is presented and published that corresponds to the real French subtransmission grid on which the proposed controller is analyzed in terms of robustness against model mismatch, constraint satisfaction, and tracking performance. It is shown that OFO controllers can help utilize the grid to its full extent, virtually reinforce it, and operate it optimally and in real-time by using the flexibility offered by renewable generators connected to distribution grids.
	\end{abstract}


    \end{frontmatter}

	\section{Introduction}
	More and more renewable power generation is installed in the grid to achieve the climate goals~\cite{world_energy_outlook_2021,renewable_energy_statistics_2022} and energy independence. In some areas, the grid's capacity is partly reached, and the uncontrolled generation of renewable power can cause overloaded lines and overvoltages.

    Grid operators have deployed different solutions to tackle this issue.
    A common approach to prevent overloaded lines is to curtail the renewable generation to a fixed maximum value, depending on the seasonal thermal ratings of the lines, without taking consumption and other generation into account. For example, these actions currently need to be implemented in the Blocaux area in France which will be used as a benchmark in this paper. This solution leads to unnecessary curtailment of renewable generation~\cite{RTE_report_virtual_reinforcement}.
	The current practice for voltage control depends on the country and ranges from manual to automatic control with large sampling times.
    The fast-changing power injections of renewable energy sources like wind and solar require higher control rates to enforce voltage and current limits in the grid.
    Because of this, manual operation strategies or automatic control with sampling times in the minutes will be incapable of safely operating a highly loaded, uncertain, and variable grid in the future. 
    Hence, increasing real-time automation of transmission grid operations is needed with control actions taken every few seconds.
	
    Real-time control does not only allow the operation of the grid under variable renewable generation, but it can also mitigate the need to physically extend the grid by virtual reinforcement through automatic control. A report by the French transmission grid operator RTE estimates possible savings of~7~billion Euros over~10~years through using real-time control of active power flows instead of building new power lines~\cite{RTE_report_virtual_reinforcement}. This opportunity comes from the higher degree of controllability of the grid, given by the flexibility of a fine network of renewable generators connected to almost every bus.

    The task of real-time efficient operation of the transmission grid can be stated as an \emph{online} optimization problem.
    In fact, at every instant in time, the optimal operation of the grid can be defined via a \ac{OPF} problem, whose parameters, however, are time-varying as they depend on the instantaneous loads, the availability of generation, etc. 

    From a control perspective, one wants to design a feedback control policy (that is, a strategy driven by real-time grid measurements) so that the state of the controlled power grid tracks the optimal state defined as the solution of a nonlinear \ac{OPF}.
    Such a real-time controller can be designed using the method known as \ac{OFO}.    
    This method was specifically developed to drive a dynamical system to the solution of an optimization problem while guaranteeing constraint satisfaction \cite{hauswirth2021optimization, bernstein2019online, lawrence2020linear, colombino2019online, bianchin2021time,simonetto2020time}. It needs minimal model information, it responds to time-varying disturbances on the systems, and it was experimentally validated in different small-scale power systems~\cite{ortmann2020experimental, ortmann2020fully,kroposki2020good}.
    
    Related earlier works also employed feedback controllers to drive a power system to an optimal operation point, but in limited specialized contexts (see~\cite{molzahn2017survey} for a review). 
    Most papers analyze the Volt/VAR problem, e.g.~\cite{bolognani2015voltage,li2022robust, qu2019optimal, liu2017distributed} and until now nearly all solutions were applied to distribution grids, e.g.~\cite{guo2023online, olives2023holistic, dominguez2023online, nowak2020measurement, picallo2022adaptive}. Recently, the publication~\cite{tang2020measurement} analyzed such algorithms for transmission grids but again limited to voltage control. For a voltage control problem in a distribution grid, the authors of~\cite{tang2020distributed} include tap changers in their control decisions.
    The application of online optimization solutions to power grids is, therefore, mostly limited to voltage control in distribution grids and, with one exception, limited to continuous inputs. Furthermore, most of them assume to have perfect model information and only consider a restricted class of actuators (e.g., only reactive power injections).
    
    Synthetic grid models are typically used to analyze the performance of these solutions. However, these models lack essential properties of real power grids e.g., locally controlled tap changers or independently operating voltage-controlled generators. Therefore, important features of these algorithms, like constraint satisfaction, tracking of time-varying conditions, and robustness to model mismatch, have not yet been analyzed on a real grid benchmark, which is essential to pave the way toward deployment. 
    
    The contributions of this work are manifold. 
    This paper presents and makes publicly available a novel subtransmission benchmark model representing a real congestion control challenge in the French transmission and subtransmission grid. The \ac{OFO} method is extended and tailored with the capability to handle discrete actuators like tap changers and it is demonstrated how such controllers can be tuned. A unified controller is designed to manage both active and reactive power injections, as well as tap changer operations. On the provided benchmark, it is demonstrated how it virtually reinforces the grid (reducing curtailment of renewable energy), and the dynamic tracking performance, the resulting constraint satisfaction, and the robustness to model mismatch.

	\section{The Unicorn 7019 Benchmark}\label{sec:benchmark}

	\begin{figure}[btp]
		\centering
		\includegraphics[width=\columnwidth]{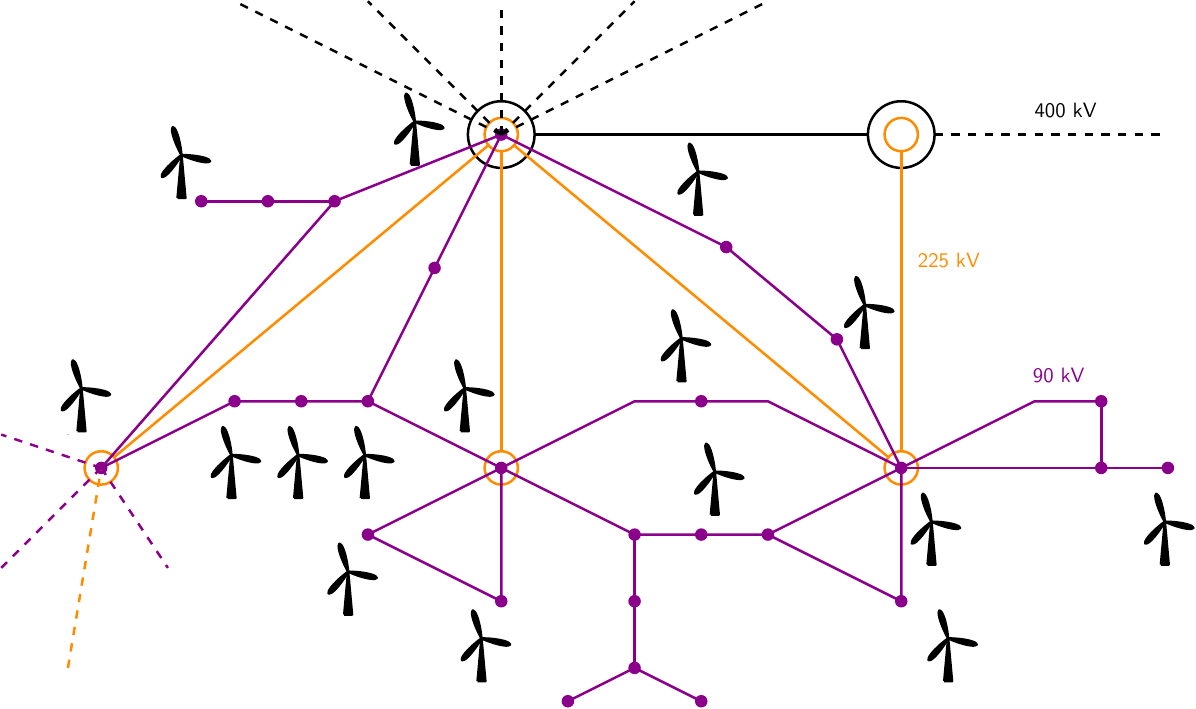}
		\caption{The Blocaux area with 31 buses, 58 branches, and 42 wind farms with a total power of 1274~MW. Connections to the rest of France are indicated with dashed lines. The tap changers are on the transformers connecting the different voltage levels.}
		\label{fig:blocaux}
	\end{figure}
    
	In this section, the Unicorn 7019 benchmark is presented, which represents the real French subtransmission grid. 
    It is implemented in MATPOWER~\cite{zimmerman2010matpower} and Matlab Simulink through the toolbox SimulinkMATPOWER~\cite{SimulinkMatpower}.
	The grid model consists of 7019 buses, 9657 branches, 1465 generators, and 907 tap changers. The steady-state behavior of the whole grid is being simulated. 
    
    Figure~\ref{fig:blocaux} shows the Blocaux area, a portion of subtransmission grid that is located in the north of France. 
    The Blocaux area consists of 31 buses, 58 branches, and there are 10 on-load tap changers on the transformers between the transmission grid (225~kV and 400~kV) and the subtransmission grid (90~kV). There are 42 wind farms with power ratings between 0.5~MW and 102~MW, and a total installed wind power of 1274~MW.

	The wind generation exceeds the power transfer capacity of the grid, and during the summer of 2021, the wind farms were curtailed at a fixed level to prevent overloaded lines. 
	The task in the benchmark is to minimize the losses and active power curtailment in the Blocaux area using the active and reactive power injections of the wind farms and the on-load tap changers, while satisfying the grid constraints, i.e., voltage magnitude limits at the buses and power flow limits on the lines.
	During the simulation the wind power produced by the wind farms is changing rapidly. Real measurements from a wind farm located close to the Blocaux area (blue line in the upper-right panel of Figure~\ref{fig:results_objective_1}) are being used. This specific wind profile is chosen because the fast rate of change makes tracking the time-varying optimum difficult and satisfying the constraints hard.

\begin{figure}[tb]
		\centering
		\includegraphics[width=\columnwidth]{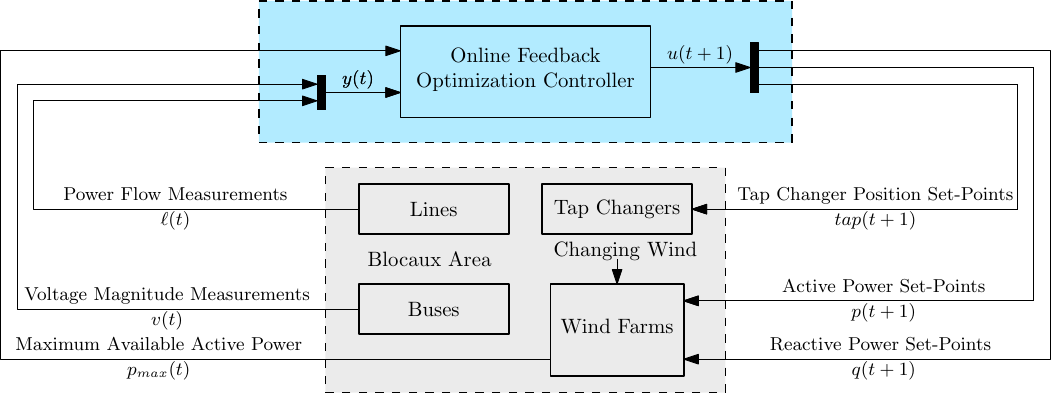}
		\caption{Block diagram of the control setup with the controller in blue and the grid in gray.}
		\label{fig:block_diagram}
	\end{figure}

	It is assumed that the active and reactive power of the 42 wind farms and the position of all 10 tap changers can be controlled. The control architecture can be seen in Figure~\ref{fig:block_diagram} and is as follows.
	Measurement devices take voltage magnitude measurements~$v$ and apparent power flow measurements~$\ell$.
    These are sent to a centralized location (a regional SCADA system) where the controller is implemented.
    The controller calculates the reactive and active power setpoints and tap changer positions.
    These are then communicated to the wind farms and tap changers. Note, that the wind farms accept continuous setpoints, whereas the tap changers are discrete actuators and only accept 33 equally spaced values between 0.9 and 1.1. It is assumed that the wind farms also communicate the currently available wind power to the controller, for example based on a wind speed measurement.

    The controller in the Blocaux area (and in any other subtransmission area) is not tasked to perform frequency control because this is done at the transmission grid level. 

    In the simulator, whenever the controller has updated its control setpoints, the power flow for the entire French grid is computed.

Afterward, all tap changers not governed by the controller — i.e., those located outside the control area — assess whether their secondary voltage is within the specified bounds.
    
    Otherwise, they switch taps, and the power flow is solved again with the updated tap ratios until no further tap changes occur. This mimics the operating behavior of the real power system.

	To showcase the controller's capabilities, two tasks are defined with which the controller is challenged.
	
	\textbf{Task 1: Optimal and Safe Curtailment} --
	minimize active power curtailment of renewable wind generation and losses while guaranteeing the satisfaction of grid constraints at all times.
	
	\textbf{Task 2: Voltage support} --
	provide voltage support as an ancillary service, keeping the voltage level of the 225~kV buses below 1.05~p.u.
	
	The benchmark is available online~\cite{gitlab}, and other researchers are invited to evaluate their control methods on it.
	

	\section{Online Feedback Optimization with Integer Constraints}\label{sec:OFO_setup}
	
	This section, provides a mathematical formulation of the control design problem, derives the \ac{OFO} controller, and comments on the necessary model information.
    
	\subsection{Input, Output, and Optimization Problem}

	The input $u=[q^\top,p^\top,tap^\top]^\top$ consists of the reactive power setpoints $q \in \mathbb{R}^{42}$ and active power setpoints $p \in \mathbb{R}^{42}$ for the 42 wind parks, and the tap positions $tap \in \mathbb{Z}^{10}$ for the 10 tap changers within the Blocaux area, for a total of 94 signals.
	The output $y=[v^\top,\ell^\top]^\top$ consists of the voltage magnitudes $v \in \mathbb{R}^{31}$ at the 31 buses and the magnitudes of the power flows $\ell \in \mathbb{R}^{58}$, for a total of 89 signals.
    The actuators that will implement our control input $u$ are limited in their capabilities and the bus voltages and line flows have lower and upper limits due to safety constraints. Therefore, the constraint sets $\mathcal{U}$ for the input and $\mathcal{Y}$ for the output are being introduced. The tap changer positions need to be an element of $\mathbb{Z}$, which is the set of integers.

	Let $d$ be the disturbance vector which contains active and reactive power consumptions and the power generation that are not controlled by the controller.
	The steady-state relation between $u$, $d$, and $y$ is determined by the power flow equations. An explicit nonlinear mapping $y=h(u,d)$ locally exists but its analytical form is generally not available~\cite{bolognani2015fast}.
    
	The following optimization problem defines, at any specific instant in time, the optimal operating point of the grid. 		    
    \begin{align}\label{eq:op_problem}
	       \begin{split}
	        	        \min_{u,y} \quad  & f(u,y) := 
                        curtailment(u)\\
	        	        \text{s.t.}\quad  &y=h(u,d)\\
	        	        &u\in \mathcal{U}, \quad
	        	        y \in \mathcal{Y}\\
	        	        &u_i \in \mathbb{Z} \quad \forall \; \text{discrete input} \; i 
	       \end{split}
	  \end{align}
	At a local optimum $(u^\star,y^\star)$, the system satisfies the constraints on $u$ and $y$, which means that it is a feasible operating point. Furthermore, the cost function is locally minimized, which corresponds to minimizing 
    the active power curtailment that is necessary to satisfy line limits.
    Implicitly, \eqref{eq:op_problem} defines a time-varying reference that the grid needs to track, as opposed as standard control design tasks where the reference to be tracked is set a-priori, or fixed. 
	
\subsection{\ac{OFO} Controller}

In order to design a feedback controller that can track the time-varying solution of \eqref{eq:op_problem}, the work from~\cite{haberle2020non} is extended and the capability to handle discrete inputs and rate limits is added.
The resulting \ac{OFO} controller has an integral form, where inputs are iteratively updated as
	\begin{equation} \label{eq:feedbackupdate} 
    u(k+1) = u(k) + \sigma(u(k),d(k),y_m(k)),
\end{equation}
where \textcolor{black}{$y_m(k)$ is a measurement of $y$ and}
\begin{align}\label{eq:projection_MIQP}
\begin{split}
    \sigma(u,d,y_m) = \arg\min_{w\in\bbR^r} \, &\| w + G^{-1} H(u,d)^\top \nabla f (u,y_m) \|^2_{G}
    \\ 
    \text{subject to} \  &A (u+w)\leq b \\ & C (y_m+\nabla_u h(u,d)w)\leq c\\
    & w \in \mathcal{W}\\
    & w_i \in \mathbb{Z} \quad \forall \; \text{discrete input} \; i,
    \end{split}
\end{align}
with $H(u,d)^\top:=\begin{bmatrix}I &\nabla_u h(u,d)^\top\end{bmatrix}$ where $\nabla_u h(u,d)$ is the sensitivity of the output with respect to the input, $I$ is the identity matrix, and $G$ is the tuning matrix.
The gradient of the cost function from~\eqref{eq:op_problem} is $\nabla f(u,y)$.
The set $\mathcal{W}$ can be used to enforce constraints on the rate of change of the input $u$\textcolor{black}{(e.g., power generation ramping limits)}.

\textcolor{black}{The matrices $A$ and $C$ and vectors $b$ and $c$ represent $\mathcal{U}$ and $\mathcal{Y}$, respectively, as intersections of halfplanes. 
They are therefore time-invariant and precomputed if $\mathcal{U}$ and $\mathcal{Y}$ are polytopes or if they can be approximated with polytopes, which is true for the vast majority of grid congestion constraints -- e.g., upper and lower limits on voltages and powers, apparent power limits, line flow limits.
If an exact linearization of the constraints is needed around the current current operating point, then standard constraint linearization methods can be employed \cite[Chapter 18]{nocedal1999}. For example, if the set $\mathcal U$ is described by the fully general inequality $g(u)\le 0$, then the Taylor expansion $g(u+w) \approx g(u)+\nabla g(u)^\top w$ immediately yields the desired linearization to be employed in~\eqref{eq:projection_MIQP}:
\[
    \underbrace{\nabla g(u)^\top}_{A} (u+w) \le \underbrace{-\nabla g(u)^\top u - g(u)}_{b}.
\]
}

Problem~\eqref{eq:projection_MIQP} is a \ac{MIQP} that needs to be solved at every time step.  It is important to notice that \eqref{eq:projection_MIQP} does not contain the nonlinear power flow equations that model the grid, and it is therefore significantly easier to solve than an \ac{OPF} like \eqref{eq:op_problem}.
The key idea of the \ac{OFO} controller is to repeatedly solve a simpler optimization program that incorporates the real-time feedback measurements from the system, knowing that the repeated closed-loop application of these updates will drive the system to the solution of \eqref{eq:op_problem} (see \cite{hauswirth2021optimization} for a review of the formal convergence guarantees). 
\textcolor{black}{Notice that \ac{OFO} inherits the same convergence properties of the iterative optimization algorithm that it implements in closed loop on the system, in this case a projected gradient descent algorithm.
For this reason, convergence cannot be guaranteed for general integer constraints.
The integer variables that appear in the grid congestion problem emerge from the discretization of some actuators (e.g., tap changers). 
The numerical simulations will show that this seems to be a favorable setting for convergence, even in a dynamic setting.}

\subsection{Controller Tuning}
\label{ssec:tuning}
\textcolor{black}{Besides defining the optimization problem~\eqref{eq:op_problem}, the only parameter in the \ac{OFO} controller that is to be tuned consists of the tuning matrix $G$, which is used to adjust the transient behavior of the closed-loop system.
In the absence of discrete inputs, the matrix G does not have an effect on the steady-state optimizer that the closed-loop system converges to. If there are discrete inputs, the matrix G does have an influence on the stead-state optimizer that the closed-loop system converges to. However, as long as stability of the closed-loop system is achieved, constraint satisfaction is always guaranteed. Overall, the choice of $G$ affects two important features of the grid transient behavior of the grid.}

\subsubsection{Rate of change of different inputs}
\textcolor{black}{For simplicity, let $G$ be a diagonal matrix $G=diag(g_1,g_2,...g_i,...g_p)$.} The bigger an entry $g_i$ of $G$, the less aggressively the corresponding input $u_i$ will be \textcolor{black}{changed at each time step} during the transient. Changing an entry $g_i$ to $g_i + \lambda$ is ultimately equivalent to adding the term $\lambda w_i^2$ to the cost function in~\eqref{eq:projection_MIQP} which adds a weight on the change of input $i$. \textcolor{black}{One can therefore regularize the rate of change of an input by increasing its corresponding value $g_i$ in the tuning matrix $G$.}

\subsubsection{Tap changer usage}
Tap changer usage comes at a cost because they slowly deteriorate with every tap change. This deterioration is irrelevant when tap changes are needed to enforce constraints. However, one might want to limit the number of tap changes used to minimize the cost function. The proposed \ac{OFO} controller conveniently provides a tuning option for this which is the matrix $G$. While $G$ does not affect the constraint enforcement it does affect how often discrete inputs are used. The higher the values in $G$ corresponding to a discrete actuator like tap changers, the less they are used. Numerical results will be shown in Section~\ref{sec:results}.

\subsection{Necessary Model Information}
\label{sec:necessaryModelInformation}
\ac{OFO} is a mostly model-free approach aside from one key piece of model information: the sensitivity $\nabla_u h(u,d)$ that describes the effect of a change in the input $u$ on the output $y$.
Note, that for example, the sensitivity of power flows (outputs) with respect to active power injections (inputs) are the well-known power-transfer-distribution-factors (PTDFs). Power system sensitivities can be computed with the implicit function theorem using the admittance matrix of the grid, the grid state, and the power flow equations~\cite{bolognani2015fast}.
The sensitivity $\nabla_u h(u,d)$ depends on both $u$ and $d$ and the system parameters, e.g. topology and line impedances. Fortunately, for many applications it can be approximated as a constant nominal parameter~\cite{ortmann2020experimental}: due to the feedback nature of \ac{OFO}, the controllers are robust against such approximations and drive the system to an operating point $(u,y)$ that satisfies the constraints and enables the safe operation of the power grid. The suboptimality of this operating point can be bounded~\cite{colombino2019towards}. 
Additionally, there exist methods to learn the sensitivity online from measurements~\cite{picallo2022adaptive, dominguez2023online, nowak2020measurement}, 
\textcolor{black}{methods to robustify the \ac{OFO} iterations against model mismatch~\cite{chan2025regularization},} and \ac{OFO} controllers that do not use any sensitivity but instead rely on zeroth order evaluation of the cost~\cite{he2022model}.
	
\subsection{Comparison between Offline Optimization and \ac{OFO}}

\begin{figure}[t!]
  \centering
  \includegraphics[width=0.6\columnwidth]{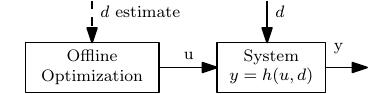}\\[3mm]
\label{fig:block_diagram_offline_opt}
  \includegraphics[width=0.6\columnwidth]{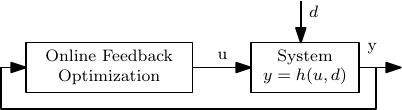}
  \caption{Offline vs. Online Feedback Optimization.}
\label{fig:block_diagram_OFO}
     \label{fig:block_diagram_offline_vs_OFO}
 \end{figure}  

As the \ac{OFO} controller ultimately drives the system to the solution of the optimization problem \eqref{eq:op_problem}, it is natural to consider whether \eqref{eq:op_problem} could be solved in the first place, and its solution used as set-points for the grid actuators. This potential approach will be referred to as Offline Optimization, as no online information (measurements) is incorporated in the decision.
    
The key difference between Offline Optimization and \ac{OFO} is that the former makes decisions based on a model and on an estimate of the disturbances, whereas the latter makes decisions based on repeated measurements, see Figure~\ref{fig:block_diagram_offline_vs_OFO}.
With perfect model information, both approaches converge to a locally optimal operation point $(u^\star,y^\star)$. 
For that, Offline Optimization needs to know the sensitivity $\nabla_u h(u,d)$ (as it is typically employed in the numerical optimization procedure), the model $h(u,d)$, and disturbance $d$. \ac{OFO}, instead, only requires the sensitivity $\nabla_u h(u,d)$, because it substitutes the model evaluation with a measurement from the plant.
In the presence of model mismatch, both Offline Optimization and \ac{OFO} necessarily converge to suboptimal points.
When using Offline Optimization, this suboptimality can yield constraint violations that may not be tolerable by the grid. \ac{OFO} controllers, instead, guarantee constraint satisfaction at steady-state thanks to including repeated measurements of the constrained variables as feedback from the system.
For a more in-depth comparison between the offline solution of the \ac{OPF} and \ac{OFO}, see Table~\ref{tab:comparision_optimization}.
For a comparison with extremum seeking, model predictive control, and modifier adaptation, see~\cite{hauswirth2021optimization}.

\begin{table*}[tb]
	    \centering
        \small
	    \begin{adjustbox}{max width=\textwidth}
	    \begin{tabular}{@{}>{\raggedright}p{0.2\textwidth}>{\raggedright}p{0.4\textwidth}>{\raggedright\arraybackslash}p{0.4\textwidth}@{}}
	        &Offline Optimization & Online Feedback Optimization \\
	        \toprule
	        Without model mismatch & Locally optimal solution & Same locally optimal solution\\
	        \midrule
	        With model mismatch & Constraint violations or suboptimal & Constraints are satisfied, converges to the best achievable solution given the model mismatch\\
         \midrule
         Computation & Solving a computationally intense non-convex problem & Calculating an easier update step that is computationally lighter
         \\
         \midrule
         Tracking behavior & No tracking (only adjusts the inputs when a new estimate of $d$ becomes available) & Tracks the time-varying optimal solution using measurements\\
         \midrule
         Communication infrastructure & Sending setpoints & Sending setpoints \& receiving measurements\\
         \midrule
         Control strategy & Feedforward & Feedback\\
         \midrule
         Decision basis & Model-based & Measurement-based\\
         \midrule
         Necessary information & Model $h(u,d)$, disturbance $d$, and sensitivity $\nabla_uh(u,d)$ & Sensitivity $\nabla_uh(u,d)$\\
         \bottomrule
	    \end{tabular}
	    \end{adjustbox}
	    \caption{Comparison of Offline Optimization and Online Feedback Optimization}
	    \label{tab:comparision_optimization}
	\end{table*}

\section{Controller Design for the Blocaux Area}\label{sec:OFO_design}

In this section, the newly proposed algorithm is presented and tailored to the Blocaux area.

\subsection{Constraint Sets and Integer Constraints}

The constraint set of the input is $\mathcal{U}=\{u\in\mathbb{R}^{94}\,|\, u_{min} \leq u \leq u_{max}\}$, where $u_{min}=[q_{min}^\top,p_{min}^\top,tap_{min}^\top]^\top$ and $u_{max}=[q_{max}^\top,p_{max}^\top,tap_{max}^\top]^\top$. 
The limits on $q$ and $tap$ are the real French grid limits and can be found in the MATPOWER case file of the benchmark~\cite{gitlab}. The lower limit for $p$ is $p_{min}=0$ because the wind farms cannot consume power.
The upper limit $p_{max}$ depends on the wind, and it is assumed that the wind farms provide this information to the controller, see the block diagram in Figure~\ref{fig:block_diagram}. Therefore, the constraints on the active power are different at every time step. An equivalent definition of the constraint set $\mathcal{U}$ is $\mathcal{U}=\{u\in \mathbb{R}^{94} \, | \, Au \leq b \}$ with $A =[I_{94}, -I_{94}]^\top \in \mathbb{R}^{188\times94}$, where $I_{94}$ is the identity matrix of size 94, and $b=[u_{max}^\top, u_{min}^\top]^\top \in \mathbb{R}^{\textcolor{black}{188}}$. This matrix $A$ and vector $b$ are used in the controller in~\eqref{eq:projection_MIQP}. 
\textcolor{black}{The specifications for the congestion control problem in the Blocaux area do not include rate limits on the actuation signals.}

The constraint set of the output is $\mathcal{Y}=\{y\in\mathbb{R}^{89}\,|\, y_{min} \leq y \leq y_{max}\}$, where $y_{min}=[v_{min}^\top,\ell_{min}^\top]^\top$ and $y_{max}=[v_{max}^\top,\ell_{max}^\top]^\top$. The limits on $v$ and $\ell$ are the real French grid limits for these buses and lines and can be found in the MATPOWER case file of the benchmark~\cite{gitlab}. An equal definition of the constraint set $\mathcal{Y}$ is $\mathcal{Y}=\{y\in \mathbb{R}^{89} \, | \, Cy \leq c \}$ with $C =[I_{89}, -I_{89}]^\top \in \mathbb{R}^{178\times89}$, where $I_{89}$ is the identity matrix of size 89, and $c=[y_{max}^\top,y_{min}^\top]^\top \in \mathbb{R}^{178}$.
This matrix $C$ and vector $c$ are used in the controller in~\eqref{eq:projection_MIQP}.
The tap changers can take 33 discrete positions, and the change of a tap position has to be an integer. 
\textcolor{black}{When calculating the solution of~\eqref{eq:projection_MIQP}, the solver must be instructed to enforce the integer constraint on the corresponding values of the decision variable $w$. This is done in line 4 of the file "OFO\_with\_integer.m" which can be found on gitlab~\cite{gitlab}.}

\subsection{Sensitivities}
The \ac{OFO} controller needs the sensitivity $\nabla_u h(u,d)$ to drive the system to the optimum. Namely, these are the sensitivities of the outputs $v$ and $\ell$ with respect to the inputs $q$, $p$, and $tap$. The sensitivity is calculated once initially for a grid state with high power generation and the true time-varying sensitivity is approximated with this fixed sensitivity. 
\textcolor{black}{The effect of this choice is assessed in detail in Section~\ref{sec:effectOfApproximations}, where it is compared with the case in which exact sensitivities are computed at every grid condition.}

\subsection{Cost Function}

\textcolor{black}{The cost function encodes the control goal, i.e., to minimize curtailment.
In order to guarantee well-posedness of the decision problem, and in particular uniqueness of the minimizer of \eqref{eq:op_problem}, it is necessary to regularize the problem by penalizing the control effort in the cost function, even slightly.
While the effect on the minimizer is negligible, it promotes a parsimonious use of the actuation inputs and it prevents chattering behaviors in dynamic settings.}

{\color{black}To design an appropriate regularization of the problem at hand, two additional terms besides the cost of curtailment are included:
\[
f(u,y) = - sum(p) + w_q \|q\|^2 + {w}_{tap}^\top tap.
\]

The term $- sum(p)$, when minimized, promoted maximal power production (minimal curtailment).

The term $w_q \|q\|^2$ can be interpreted as an approximation of the additional losses due to reactive power flows on the medium voltage lines connecting the wind farms to the high voltage grid. The scalar weight has been set to $w_q=0.0026$ based on the analysis of typical medium voltage feeders in the area.

The term ${w}_{tap}^\top tap$ is a linear approximation of the effect of the tap position on the grid losses (which are reduced for higher tap positions). ${w}_{tap}$ is a vector of size 10 and corresponds to the derivative of the losses with respect to the tap changer positions. It is calculated once numerically for a grid state with high power generation and kept constant throughout the simulation. The values are
\[w_{tap}=\begin{bmatrix}
    -9.88 &-0.00 &-0.00 &-0.00 &-10.8 &-10.3 &-2.21 &-2.20 &-9.53 &-9.52
\end{bmatrix}^\top\]
}

Note, that for the implementation of the controller only the gradient of the cost function $\nabla f(u,y)$ is needed, see~\eqref{eq:projection_MIQP}. 
The gradient of our cost function is
\textcolor{black}{\begin{equation*}
\nabla f(u,y) =
\begin{bmatrix}
     \nabla_u f(u,y)\\
     \cmidrule(lr){1-1}
     \nabla_y f(u,y)
\end{bmatrix}\approx 
\begin{bmatrix}
     2 {w}_q q\\
     -1_{42}\\
     {w}_{tap}\\
     \cmidrule(lr){1-1}
     0_{89}
\end{bmatrix},
\end{equation*}
where $1_{42}$ is a vector of ones and $0_{89}$ is a vector of zeros of dimension 42 and 89, respectively.}

\subsection{Tuning Parameters and Sampling Time}
The sampling time of the controller is 10~seconds.
To solve the \ac{MIQP} in \eqref{eq:projection_MIQP} the Yalmip toolbox~\cite{lofberg2004yalmip} with the solver Gurobi is used. Solving the \ac{MIQP} takes less than 40~milliseconds on a standard notebook.
The tuning parameter is a diagonal matrix $G = \diag(0.1 \cdot I_{42}, 0.2 \cdot I_{42}, 2500 \cdot I_{10})$. The high value of 2500 for the tap changers ensures that they are not overused. The small values for $p$ and $q$ ensure fast algorithm convergence.

\section{Results}\label{sec:results}
	
In this section, the Unicorn 7019 benchmark is used to validate the \ac{OFO} controller and showcase its performance. Simulations for both tasks (\textbf{Optimal and Safe Curtailment} and \textbf{Voltage support}) are shown. The effect of the approximate model information is discussed by comparing its performance with the case with perfect model knowledge. 
The tracking performance of the \ac{OFO} controller is shown, i.e., how closely it tracks the time-varying solution of the \ac{OPF}, and it is compared to the current practice of curtailing injection at a fixed level. Finally, the influence of the tuning parameter $G$ on the tap changer behavior is presented.
 
\subsection{Task 1: Optimal and Safe Curtailment}
\label{ssec:resultscurtailment}
\begin{figure*}
	    \setlength\fwidth{20cm}
        \setlength\fheight{8cm}
	    \centering
\hspace*{-30mm}\scalebox{0.8}{\input{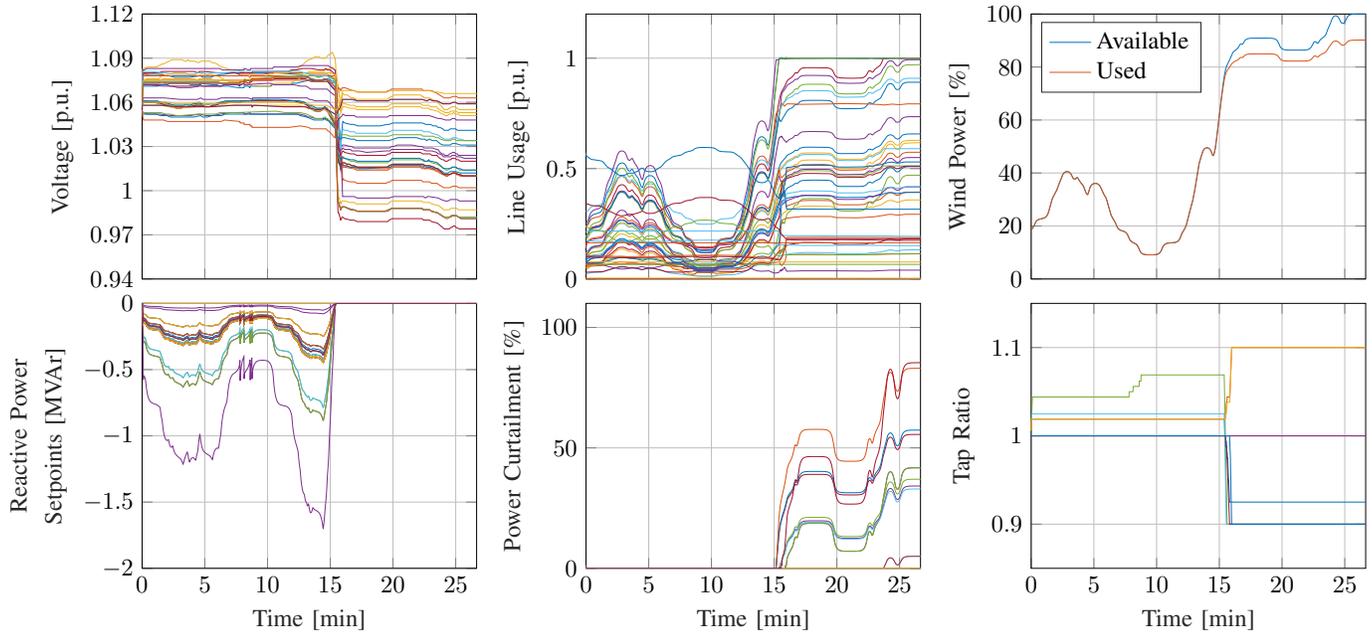}}
	    \caption{Simulation of the Unicorn benchmark under time-varying wind power (blue line in top right plot). The lower three plots show the control inputs and the upper plots show the resulting voltages, line usage, and used wind power. The voltage constraints that the controller has to enforce are the real French grid limits.}
	    \label{fig:results_objective_1}
	\end{figure*}

For the task of optimal and safe curtailment, the controller as described in Section~\ref{sec:OFO_design} is deployed. The results can be seen in Figure~\ref{fig:results_objective_1}, and the behavior can be separated into two phases. In the beginning, the wind power is low and no curtailment is necessary. Therefore, the controller aims to minimize the losses. This is done by increasing the voltage with high tap ratios. To satisfy the voltage constraints not all tap changers increase their tap positions and some reactive power is used. Note that the sharp steps in the reactive power injections (lower left panel) can arise when a tap is changed. These steps in the reactive power are sometimes needed to enforce the voltage constraints when a tap change occurs.
Around minute~15 the wind power sharply increases and surpasses the grid capacity. Curtailment becomes necessary to satisfy the current limits on some of the lines. Even though the tap changers have a small influence on the power flows, this influence is used to redirect flow to non-congested lines. This helps to minimize the necessary curtailment. The tap changes lead to voltages well within the constraints and no reactive power is needed anymore. These are highly non-trivial control actions that no human operator would have been able to derive in real-time.
While the wind power increases, the controller curtails more active power to keep the flows within the constraints. It can be seen in Figure~\ref{fig:results_objective_1} that line constraints violations are minimal and most importantly they are temporary.
Therefore, the grid is safely operated at its limit allowing a maximum of renewable wind power to be injected. At the end of the simulation 88.9\% of the wind power is injected into the grid. A closeup of the tap changer behaviour can be seen in Figure~\ref{fig:tap_changer_closeup}.
	
As explained before, the \ac{OFO} controller is tracking the locally optimal solution of the optimization problem defined in Section~\ref{sec:OFO_design}. There is a high cost on active power curtailment and therefore this control action is only used if it is necessary to satisfy line constraints or in the unlikely case that such a high amount of reactive power would need to be used to satisfy voltage limits that a lower value of the cost function could be achieved by curtailing active power.  

	\begin{figure*}
	    \setlength\fwidth{7.8cm}
        \setlength\fheight{4.5cm}
	    \centering
\scalebox{0.8}{
%
%
\definecolor{mycolor1}{rgb}{0.00000,0.44700,0.74100}%
\definecolor{mycolor2}{rgb}{0.85000,0.32500,0.09800}%
\definecolor{mycolor3}{rgb}{0.92900,0.69400,0.12500}%
\definecolor{mycolor4}{rgb}{0.49400,0.18400,0.55600}%
\definecolor{mycolor5}{rgb}{0.46600,0.67400,0.18800}%
\definecolor{mycolor6}{rgb}{0.30100,0.74500,0.93300}%
\definecolor{mycolor7}{rgb}{0.63500,0.07800,0.18400}%
\begin{tikzpicture}

\begin{axis}[%
width=0.951\fwidth,
height=\fheight,
at={(0\fwidth,0\fheight)},
scale only axis,
xmin=0.000,
xmax=26.667,
xtick={ 0,  5, 10, 15, 20, 25},
xlabel style={font=\color{white!15!black}},
xlabel={Time [min]},
ymin=0.850,
ymax=1.150,
ylabel style={font=\color{white!15!black}},
ylabel={Tap Ratio},
axis background/.style={fill=white},
xmajorgrids,
ymajorgrids
]
\addplot [color=mycolor1, forget plot]
  table[row sep=crcr]{%
0.000	1.006\\
0.150	1.006\\
0.167	1.012\\
0.317	1.012\\
0.333	1.019\\
0.483	1.019\\
0.500	1.025\\
1.483	1.025\\
1.500	1.031\\
1.650	1.031\\
1.667	1.038\\
2.150	1.038\\
2.167	1.044\\
2.483	1.044\\
2.500	1.050\\
3.483	1.050\\
3.500	1.056\\
3.650	1.056\\
3.667	1.062\\
4.483	1.062\\
4.500	1.069\\
4.650	1.069\\
4.667	1.075\\
6.317	1.075\\
6.333	1.081\\
6.483	1.081\\
6.500	1.087\\
6.817	1.087\\
6.833	1.094\\
7.817	1.094\\
7.833	1.100\\
26.667	1.100\\
};
\addplot [color=mycolor2, forget plot]
  table[row sep=crcr]{%
0.000	1.000\\
26.667	1.000\\
};
\addplot [color=mycolor3, forget plot]
  table[row sep=crcr]{%
0.000	1.000\\
26.667	1.000\\
};
\addplot [color=mycolor4, forget plot]
  table[row sep=crcr]{%
0.000	1.000\\
26.667	1.000\\
};
\addplot [color=mycolor5, forget plot]
  table[row sep=crcr]{%
0.000	1.006\\
0.150	1.006\\
0.167	1.012\\
0.317	1.012\\
0.333	1.019\\
0.483	1.019\\
0.500	1.025\\
0.650	1.025\\
0.667	1.031\\
0.817	1.031\\
0.833	1.038\\
0.983	1.038\\
1.000	1.044\\
1.150	1.044\\
1.167	1.050\\
1.317	1.050\\
1.333	1.056\\
1.483	1.056\\
1.500	1.062\\
1.650	1.062\\
1.667	1.069\\
4.650	1.069\\
4.667	1.075\\
6.650	1.075\\
6.667	1.081\\
6.983	1.081\\
7.000	1.087\\
8.650	1.087\\
8.667	1.094\\
10.317	1.094\\
10.333	1.100\\
15.317	1.100\\
15.333	1.094\\
15.483	1.094\\
15.500	1.075\\
15.650	1.075\\
15.667	1.056\\
15.817	1.056\\
15.833	1.038\\
15.983	1.038\\
16.000	1.019\\
16.150	1.019\\
16.167	1.000\\
16.317	1.000\\
16.333	0.981\\
16.483	0.981\\
16.500	0.962\\
16.650	0.962\\
16.667	0.944\\
16.817	0.944\\
16.833	0.931\\
16.983	0.931\\
17.000	0.913\\
17.150	0.913\\
17.167	0.900\\
26.667	0.900\\
};
\addplot [color=mycolor6, forget plot]
  table[row sep=crcr]{%
0.000	1.006\\
0.150	1.006\\
0.167	1.012\\
0.317	1.012\\
0.333	1.019\\
0.483	1.019\\
0.500	1.025\\
1.150	1.025\\
1.167	1.031\\
1.317	1.031\\
1.333	1.038\\
1.483	1.038\\
1.500	1.044\\
10.483	1.044\\
10.500	1.050\\
11.983	1.050\\
12.000	1.056\\
13.317	1.056\\
13.333	1.062\\
14.650	1.062\\
14.667	1.069\\
14.817	1.069\\
14.833	1.075\\
15.317	1.075\\
15.333	1.069\\
15.483	1.069\\
15.500	1.050\\
15.650	1.050\\
15.667	1.031\\
15.817	1.031\\
15.833	1.012\\
15.983	1.012\\
16.000	0.994\\
16.150	0.994\\
16.167	0.975\\
16.317	0.975\\
16.333	0.956\\
16.483	0.956\\
16.500	0.938\\
16.650	0.938\\
16.667	0.919\\
16.817	0.919\\
16.833	0.900\\
26.667	0.900\\
};
\addplot [color=mycolor7, forget plot]
  table[row sep=crcr]{%
0.000	1.000\\
15.483	1.000\\
15.500	0.988\\
15.650	0.988\\
15.667	0.975\\
15.817	0.975\\
15.833	0.962\\
15.983	0.962\\
16.000	0.950\\
16.150	0.950\\
16.167	0.938\\
16.317	0.938\\
16.333	0.931\\
16.483	0.931\\
16.500	0.925\\
16.650	0.925\\
16.667	0.919\\
16.817	0.919\\
16.833	0.913\\
16.983	0.913\\
17.000	0.900\\
26.667	0.900\\
};
\addplot [color=mycolor1, forget plot]
  table[row sep=crcr]{%
0.000	1.000\\
15.483	1.000\\
15.500	0.988\\
15.650	0.988\\
15.667	0.975\\
15.817	0.975\\
15.833	0.962\\
15.983	0.962\\
16.000	0.950\\
16.150	0.950\\
16.167	0.944\\
16.317	0.944\\
16.333	0.938\\
16.483	0.938\\
16.500	0.931\\
16.650	0.931\\
16.667	0.925\\
16.817	0.925\\
16.833	0.919\\
16.983	0.919\\
17.000	0.913\\
17.150	0.913\\
17.167	0.900\\
26.667	0.900\\
};
\addplot [color=mycolor2, forget plot]
  table[row sep=crcr]{%
0.000	1.006\\
0.150	1.006\\
0.167	1.012\\
0.317	1.012\\
0.333	1.019\\
4.650	1.019\\
4.667	1.025\\
7.983	1.025\\
8.000	1.031\\
8.817	1.031\\
8.833	1.038\\
11.817	1.038\\
11.833	1.044\\
12.483	1.044\\
12.500	1.050\\
12.983	1.050\\
13.000	1.056\\
13.150	1.056\\
13.167	1.062\\
14.650	1.062\\
14.667	1.069\\
14.817	1.069\\
14.833	1.075\\
15.483	1.075\\
15.500	1.081\\
15.650	1.081\\
15.667	1.087\\
15.817	1.087\\
15.833	1.094\\
15.983	1.094\\
16.000	1.100\\
26.667	1.100\\
};
\addplot [color=mycolor3, forget plot]
  table[row sep=crcr]{%
0.000	1.006\\
0.150	1.006\\
0.167	1.012\\
0.317	1.012\\
0.333	1.019\\
4.650	1.019\\
4.667	1.025\\
14.650	1.025\\
14.667	1.031\\
14.817	1.031\\
14.833	1.038\\
15.483	1.038\\
15.500	1.044\\
15.650	1.044\\
15.667	1.050\\
15.817	1.050\\
15.833	1.056\\
15.983	1.056\\
16.000	1.062\\
16.150	1.062\\
16.167	1.069\\
16.317	1.069\\
16.333	1.075\\
16.483	1.075\\
16.500	1.081\\
16.650	1.081\\
16.667	1.087\\
16.817	1.087\\
16.833	1.094\\
16.983	1.094\\
17.000	1.100\\
26.667	1.100\\
};
\end{axis}
\end{tikzpicture}%
}
	    \caption{Closeup of the tap changer behavior between minutes 14 and 18.}
	    \label{fig:tap_changer_closeup}
	\end{figure*}
    
	\subsection{Task 2: Voltage support}
\begin{figure*}
	    \setlength\fwidth{20cm}
        \setlength\fheight{8cm}
	    \centering
\hspace*{-30mm}\scalebox{0.8}{\input{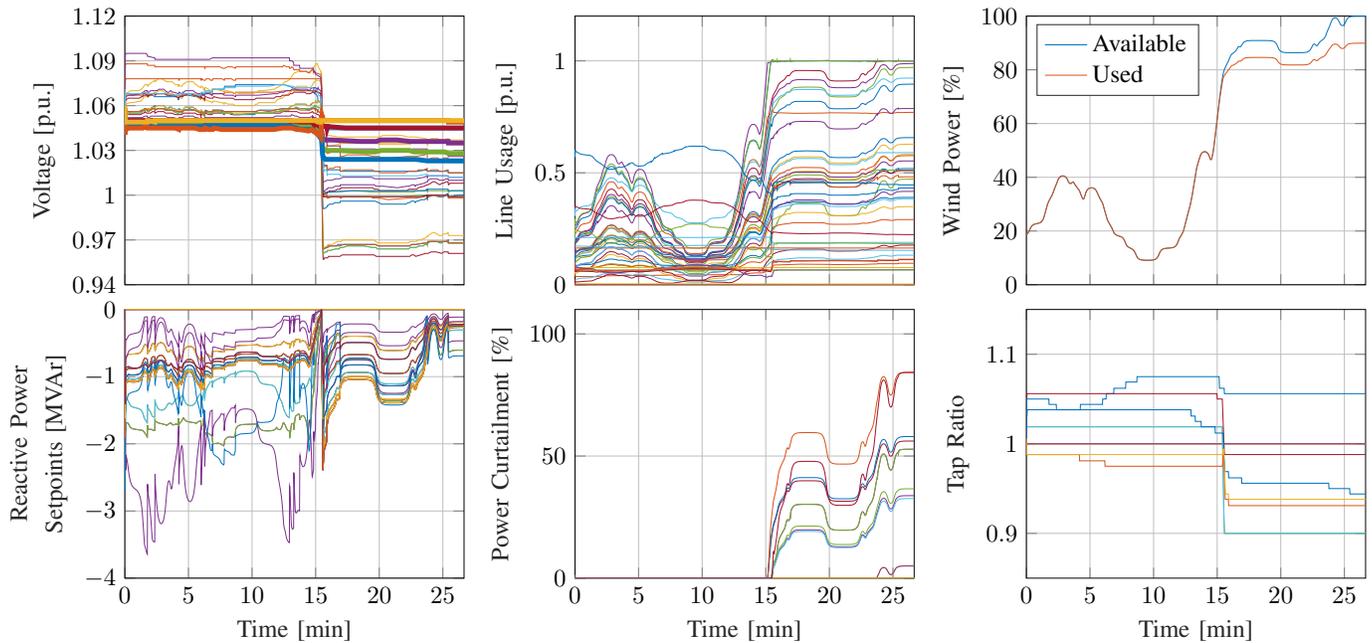}}
	    \caption{Simulation of the Unicorn benchmark under time-varying wind power (blue line in top right plot). The lower three plots show the control inputs and the upper plots show the resulting voltages, line usage, and used wind power. The voltage constraints of the 225~kV buses were set to 1.05~p.u. Those buses are indicated with a thicker line. The controller is enforcing these limits.}
	    \label{fig:results_objective_2}
	\end{figure*}

	In this task the subtransmission grid has to provide voltage support to the transmission grid as an ancillary service. The goal is to keep the voltages at the 225~kV buses below 1.05 p.u. This can easily be incorporated into our controller by changing the upper voltage constraint of these buses to 1.05. This example shows the versatility of our framework, i.e., defining the control goals through an optimization problem. The results of the simulation can be seen in Figure~\ref{fig:results_objective_2}. Similar to the previous task the behavior is split into two phases. In the beginning, the losses are minimized and after minute~15 the focus lies on optimizing curtailment as it becomes the predominant part of the cost. The main difference to the previous task is in the usage of the tap changers and reactive power to satisfy the tighter voltage constraints on the 225~kV buses. For that, more reactive power is used and the tap changers are actuated more.
    After minute~15 the tap changers behave differently than in the previous task because of the tighter voltage constraints. This leads to a slightly higher curtailment and reactive power is still needed to enforce the voltage constraints. At the end of the simulation 88.6\% of the wind power can be injected into the grid.
    
 \subsection{The Effect of the Approximations}
 \label{sec:effectOfApproximations}

\begin{figure*}
	    \setlength\fwidth{20cm}
        \setlength\fheight{8cm}
	    \centering
\hspace*{-30mm}\scalebox{0.8}{\input{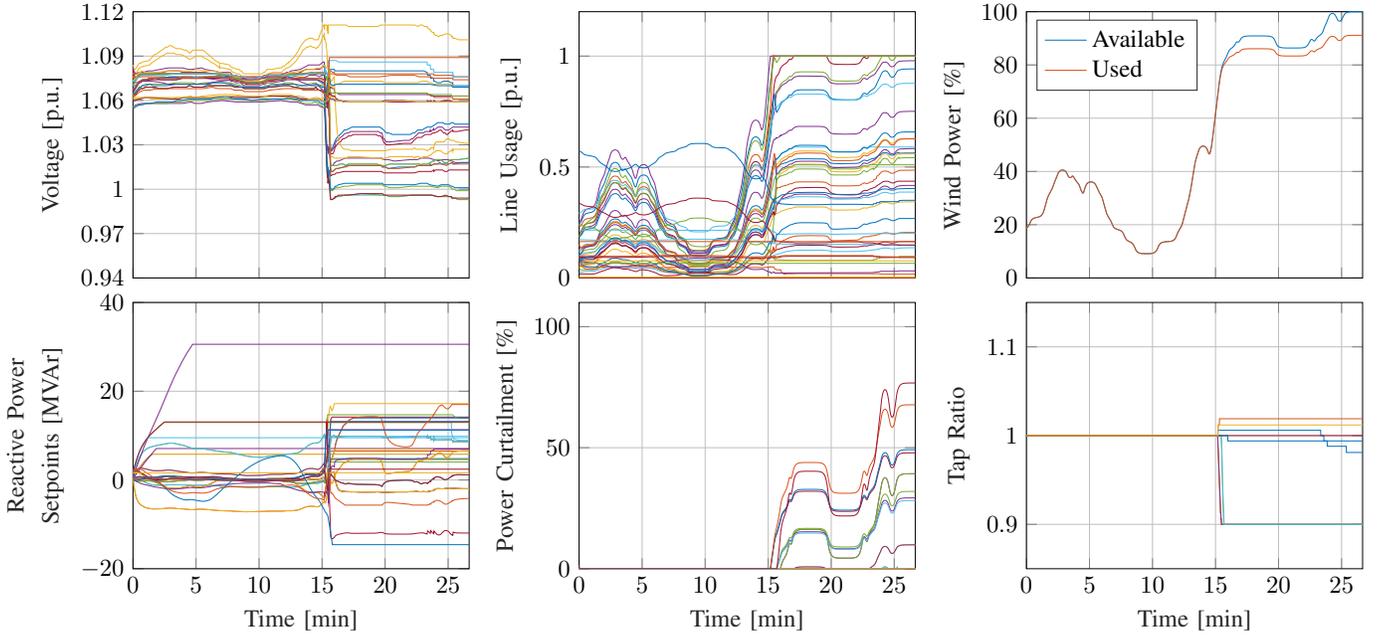}}
	    \caption{Simulation of the Unicorn benchmark under time-varying wind power (blue line in top right plot). The lower three plots show the control inputs and the upper plots show the resulting voltages, line usage, and used wind power. In this simulation, the controller has perfect knowledge of the sensitivity $\nabla_u h(u,d)$ and the derivative $\nabla f(u,y)$.}
	    \label{fig:optimality}
     
	\end{figure*}

As discussed in Section~\ref{sec:OFO_design}, the sensitivity $\nabla_u h(u,d)$ and the derivative of the cost function $\nabla f(u,y)$ are not perfectly known for the benchmark under study, which is always the case in a practical application. This imperfection affects the optimality of the time-varying state the controller is tracking.
To analyze the sub-optimality the \ac{OFO} controller for \emph{Optimal and Safe Curtailment} is run again but with perfect information about the sensitivity and derivative of the cost function at every time step.
The main difference is in the usage of reactive power and the tap changers, see Figure~\ref{fig:optimality}. The active power curtailment however is very similar. This is because the approximation of the derivative of the cost function with respect to curtailment is highly accurate as well as the sensitivity of the line flows with respect to the curtailment (those are PTDFs). Note, that the effectiveness of using a constant sensitivity matrix is not always guaranteed. Overall, 91.5\% of the available wind power can be used, which is slightly higher than what was achieved with approximate modeling information (88.9\%, see Section~\ref{ssec:resultscurtailment}).
This is mostly due to using the reactive power capabilities to reduce reactive power flows on the saturated lines and therefore freeing up capacity for active power flows. Also, the reactive power capabilities are used to reduce reactive power flows in general to lower the losses in the grid. Note, that the effectiveness of using a constant sensitivity matrix is not always guaranteed.
Again, it is highlighted that the suboptimality under model mismatch does not arise from the method but is fundamental to the problem of making decisions based on inaccurate information.
Nevertheless, \ac{OFO} guarantees constraint satisfaction in steady-state.

	\subsection{Tracking Performance and Comparison to Ground Truth}
\begin{figure}
        \setlength\fwidth{7.8cm}
        \setlength\fheight{4.5cm}
	    \centering
	\scalebox{0.8}{\input{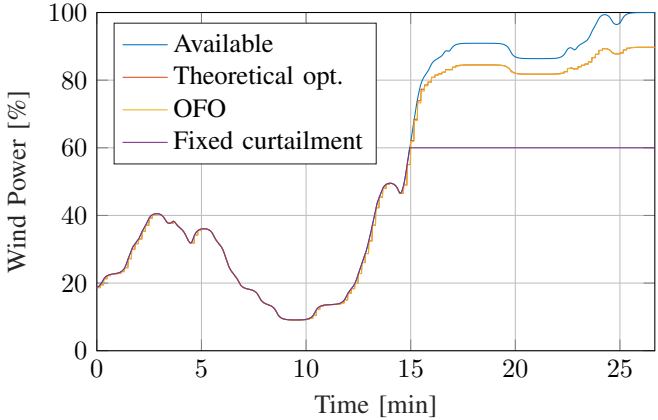}}
	    \caption{Comparison of the available wind power, the wind power that can be theoretically used, fixed curtailment at 60\%, and the wind power that the \ac{OFO} controller can use.}
	    \label{fig:tracking}
	\end{figure}

Changes in the consumption or the production affect the optimal solution of the optimization problem and an \ac{OFO} controller is constantly trying to track this time-varying local optimum. To analyze the tracking performance the time-varying optimal solution is computed with an \ac{OPF} solver that has perfect model knowledge. For a fair comparison, also the \ac{OFO} controller is provided with perfect model knowledge. The task \emph{Optimal and Safe Curtailment} is solved with both approaches and the wind power that they can use is plotted. To be able to run the MATPOWER OPF solver the tap changers in the Blocaux area blocked at~1. To have a fair comparison the tap changers are also blocked at~1 for the \ac{OFO} controller. The result can be seen in Figure~\ref{fig:tracking}, and show that the optimality of the \ac{OFO} controller is practically identical to the theoretical optimum that was computed with an omniscient and instantaneous OPF solver. The figure also shows a comparison with the state-of-the-art which is curtailing the wind farms at a fixed level. Here~60\% is chosen which already leads to one line being loaded at close to 90\% meaning that the remaining headroom for potential lower consumption in the Blocaux and therefore higher line loads is small. From the figure it becomes apparent that at full wind output, the \ac{OFO} controller allows the wind farms to inject 50~\% more power than the current industrial practice. The reason behind this large number is that not all lines become saturated and therefore only a few wind parks need to be curtailed. Due to the feedback from the grid, the \ac{OFO} controller can optimally solve this non-trivial task of deciding which wind park has to be curtailed by which value at what time while making sure all voltages and lines are within their limits.

\subsection{Tap changer behavior}
	
\begin{figure}
	   \centering
	     \setlength\fwidth{7.8cm}
        \setlength\fheight{7cm}
	    \scalebox{0.8}{\input{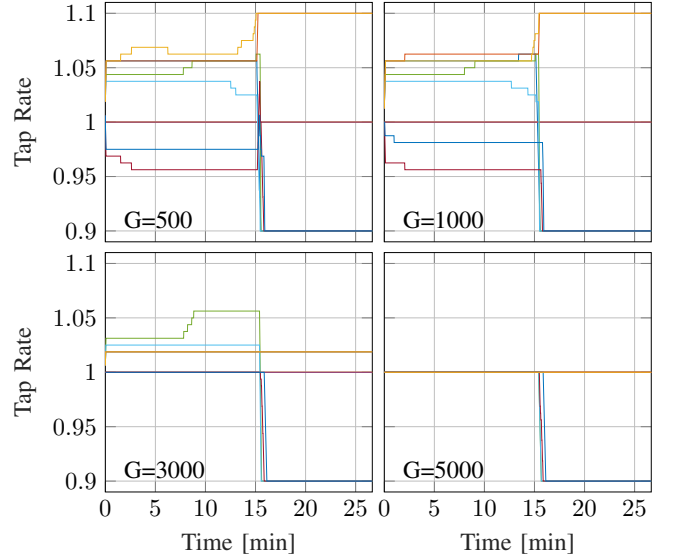}}
	    \caption{Behavior of the tap changers for different values of the entries of G corresponding to the tap changers.}
	    \label{fig:tap_changer_behavior}
	\end{figure}

	As discussed in Section~\ref{ssec:tuning}, the tuning matrix $G$ influences the tap changer behavior. To illustrate this phenomenon, the use of the \ac{OFO} controller for \emph{Optimal and Safe Curtailment} is simulated for different values of the entries of $G$ corresponding to tap changers. The results for different values can be seen in Figure~\ref{fig:tap_changer_behavior}. With a value of~500, the tap changers are heavily used and might deteriorate quickly. For larger values, their usage decreases and with the value~5000 they are only used once the grid reaches its capacity limit. With a much larger value than~5000, the tap changers would not be used at all unless needed to enforce operational constraints, e.g., voltage limits.

	\section{Conclusion}\label{sec:conclusion}
	By extending \ac{OFO} controllers with the capability to handle discrete actuators, it is possible to design a real-time controller for a real subtransmission grid benchmark that uses a diverse and complex range of actuators, i.e., active and reactive power capabilities and on-load tap changers. 
    This controller operates the grid safely at the locally optimal operation point and tracks the local optimum when it changes over time. 
    It was shown that this tracking is highly precise and not sensitive to model mismatch. This leads to the \ac{OFO} controller being able to extract 50\% more wind power in the real-world benchmark than the current industrial practice.
    Furthermore, it was shown that specifying control objectives in an optimization problem is versatile and allows us to easily include grid ancillary services like voltage control. Most importantly, the simulations on the real benchmark show that \ac{OFO} can be a powerful tool to safely operate the subtransmission grid in real-time, under model mismatch, and with limited model information. A remaining open question is how several independently controlled subtransmission grids would interact with each other. \textcolor{black}{A game-theoretic analysis of the multi-area setting predicts that the resulting coupled behavior would converge to an equilibrium condition that may not correspond to the globally optimal state of the grid \cite{belgioioso2023multiarea}. However, a quantitative assesement of this suboptimality is still lacking.}
 
\bibliographystyle{IEEEtran}
\bibliography{IEEEabrv,biblio}

\begin{thebibliography}{10}
\providecommand{\url}[1]{#1}
\csname url@samestyle\endcsname
\providecommand{\newblock}{\relax}
\providecommand{\bibinfo}[2]{#2}
\providecommand{\BIBentrySTDinterwordspacing}{\spaceskip=0pt\relax}
\providecommand{\BIBentryALTinterwordstretchfactor}{4}
\providecommand{\BIBentryALTinterwordspacing}{\spaceskip=\fontdimen2\font plus
\BIBentryALTinterwordstretchfactor\fontdimen3\font minus
  \fontdimen4\font\relax}
\providecommand{\BIBforeignlanguage}[2]{{%
\expandafter\ifx\csname l@#1\endcsname\relax
\typeout{** WARNING: IEEEtran.bst: No hyphenation pattern has been}%
\typeout{** loaded for the language `#1'. Using the pattern for}%
\typeout{** the default language instead.}%
\else
\language=\csname l@#1\endcsname
\fi
#2}}
\providecommand{\BIBdecl}{\relax}
\BIBdecl

\bibitem{world_energy_outlook_2021}
\BIBentryALTinterwordspacing
``World energy outlook 2021,'' IEA, Tech. Rep., 2021. [Online]. Available:
  \url{https://www.iea.org/reports/world-energy-outlook-2021}
\BIBentrySTDinterwordspacing

\bibitem{renewable_energy_statistics_2022}
``Renewable energy statistics 2022,'' The International Renewable Energy
  Agency, Tech. Rep., 2022.

\bibitem{RTE_report_virtual_reinforcement}
\BIBentryALTinterwordspacing
``Schéma décennal de développement du réseau (sddr) 2019 - rapport
  complet,'' RTE Réseau de transport d’électricité, Tech. Rep., 2019.
  [Online]. Available:
  \url{https://www.rte-france.com/analyses-tendances-et-prospectives/le-schema-decennal-de-developpement-du-reseau#Documents}
\BIBentrySTDinterwordspacing

\bibitem{hauswirth2021optimization}
A.~Hauswirth, Z.~He, S.~Bolognani, G.~Hug, and F.~D{\"o}rfler, ``Optimization
  algorithms as robust feedback controllers,'' \emph{Annual Reviews in
  Control}, vol.~57, p. 100941, 2024.

\bibitem{bernstein2019online}
A.~Bernstein, E.~Dall'Anese, and A.~Simonetto, ``Online primal-dual methods
  with measurement feedback for time-varying convex optimization,'' \emph{IEEE
  Trans. Signal Process.}, vol.~67, no.~8, pp. 1978--1991, 2019.

\bibitem{lawrence2020linear}
L.~S. Lawrence, J.~W. Simpson-Porco, and E.~Mallada, ``Linear-convex optimal
  steady-state control,'' \emph{IEEE Transactions on Automatic Control},
  vol.~66, no.~11, pp. 5377--5384, 2020.

\bibitem{colombino2019online}
M.~Colombino, E.~Dall’Anese, and A.~Bernstein, ``Online optimization as a
  feedback controller: Stability and tracking,'' \emph{IEEE Transactions on
  Control of Network Systems}, vol.~7, no.~1, pp. 422--432, 2019.

\bibitem{bianchin2021time}
G.~Bianchin, J.~Cortes, J.~I. Poveda, and E.~Dall'Anese, ``Time-varying
  optimization of lti systems via projected primal-dual gradient flows,''
  \emph{IEEE Transactions on Control of Network Systems}, 2021.

\bibitem{simonetto2020time}
A.~Simonetto, E.~Dall'Anese, S.~Paternain, G.~Leus, and G.~B. Giannakis,
  ``Time-varying convex optimization: Time-structured algorithms and
  applications,'' \emph{Proceedings of the IEEE}, vol. 108, no.~11, pp.
  2032--2048, 2020.

\bibitem{ortmann2020experimental}
L.~Ortmann, A.~Hauswirth, I.~Caduff, F.~D{\"o}rfler, and S.~Bolognani,
  ``Experimental validation of feedback optimization in power distribution
  grids,'' \emph{Electric Power Systems Research}, vol. 189, p. 106782, 2020.

\bibitem{ortmann2020fully}
L.~Ortmann, A.~Prostejovsky, K.~Heussen, and S.~Bolognani, ``Fully distributed
  peer-to-peer optimal voltage control with minimal model requirements,''
  \emph{Electric Power Systems Research}, vol. 189, p. 106717, 2020.

\bibitem{kroposki2020good}
B.~Kroposki, A.~Bernstein, J.~King, and F.~Ding, ``Good grids make good
  neighbors,'' \emph{IEEE Spectrum}, vol.~57, no. NREL/JA-5D00-78521, 2020.

\bibitem{molzahn2017survey}
D.~K. Molzahn, F.~D{\"o}rfler, H.~Sandberg, S.~H. Low, S.~Chakrabarti,
  R.~Baldick, and J.~Lavaei, ``A survey of distributed optimization and control
  algorithms for electric power systems,'' \emph{IEEE Transactions on Smart
  Grid}, vol.~8, no.~6, pp. 2941--2962, 2017.

\bibitem{bolognani2015voltage}
S.~Bolognani, R.~Carli, G.~Cavraro, and S.~Zampieri, ``Distributed reactive
  power feedback control for voltage regulation and loss minimization,''
  \emph{IEEE Transactions on Automatic Control}, vol.~60, no.~4, pp. 966--981,
  2015.

\bibitem{li2022robust}
S.~Li, W.~Wu, and Y.~Lin, ``Robust data-driven and fully distributed volt/var
  control for active distribution networks with multiple virtual power
  plants,'' \emph{IEEE Transactions on Smart Grid}, vol.~13, no.~4, pp.
  2627--2638, 2022.

\bibitem{qu2019optimal}
G.~Qu and N.~Li, ``Optimal distributed feedback voltage control under limited
  reactive power,'' \emph{IEEE Transactions on Power Systems}, vol.~35, no.~1,
  pp. 315--331, 2019.

\bibitem{liu2017distributed}
H.~J. Liu, W.~Shi, and H.~Zhu, ``Distributed voltage control in distribution
  networks: Online and robust implementations,'' \emph{IEEE Transactions on
  Smart Grid}, vol.~9, no.~6, pp. 6106--6117, 2017.

\bibitem{guo2023online}
Y.~Guo, X.~Zhou, C.~Zhao, L.~Chen, G.~Hug, and T.~H. Summers, ``An online joint
  optimization--estimation architecture for distribution networks,'' \emph{IEEE
  Transactions on Control Systems Technology}, 2023.

\bibitem{olives2023holistic}
J.~C. Olives-Camps, {\'A}.~R. del Nozal, J.~M. Mauricio, and J.~M. Maza-Ortega,
  ``A holistic model-less approach for the optimal real-time control of power
  electronics-dominated {AC} microgrids,'' \emph{Applied Energy}, vol. 335, p.
  120761, 2023.

\bibitem{dominguez2023online}
A.~D. Domínguez-García, M.~Zholbaryssov, T.~Amuda, and O.~Ajala, ``An online
  feedback optimization approach to voltage regulation in inverter-based power
  distribution networks,'' in \emph{2023 American Control Conference (ACC)},
  2023, pp. 1868--1873.

\bibitem{nowak2020measurement}
S.~Nowak, Y.~C. Chen, and L.~Wang, ``Measurement-based optimal {DER} dispatch
  with a recursively estimated sensitivity model,'' \emph{IEEE Transactions on
  Power Systems}, vol.~35, no.~6, pp. 4792--4802, 2020.

\bibitem{picallo2022adaptive}
M.~Picallo, L.~Ortmann, S.~Bolognani, and F.~D{\"o}rfler, ``Adaptive real-time
  grid operation via online feedback optimization with sensitivity
  estimation,'' \emph{Electric Power Systems Research}, vol. 212, p. 108405,
  2022.

\bibitem{tang2020measurement}
Z.~Tang, E.~Ekomwenrenren, J.~W. Simpson-Porco, E.~Farantatos, M.~Patel, and
  H.~Hooshyar, ``Measurement-based fast coordinated voltage control for
  transmission grids,'' \emph{IEEE Transactions on Power Systems}, vol.~36,
  no.~4, pp. 3416--3429, 2020.

\bibitem{tang2020distributed}
Z.~Tang, D.~J. Hill, and T.~Liu, ``Distributed coordinated reactive power
  control for voltage regulation in distribution networks,'' \emph{IEEE
  Transactions on Smart Grid}, vol.~12, no.~1, pp. 312--323, 2020.

\bibitem{zimmerman2010matpower}
R.~D. Zimmerman, C.~E. Murillo-S{\'a}nchez, and R.~J. Thomas, ``Matpower:
  Steady-state operations, planning, and analysis tools for power systems
  research and education,'' \emph{IEEE Transactions on power systems}, vol.~26,
  no.~1, pp. 12--19, 2010.

\bibitem{SimulinkMatpower}
L.~Ortmann, ``Github repository,''
  \url{https://github.com/Lukas738/SimulinkMATPOWER}, 2023.

\bibitem{gitlab}
------, ``Gitlab repository,''
  \url{https://gitlab.ost.ch/lukas.ortmann/subtransmission-grid-control-via-online-feedback-optimization},
  2025.

\bibitem{bolognani2015fast}
S.~Bolognani and F.~D{\"o}rfler, ``Fast power system analysis via implicit
  linearization of the power flow manifold,'' in \emph{53rd Annual Allerton
  Conference on Communication, Control, and Computing (Allerton)}, 2015, pp.
  402--409.

\bibitem{haberle2020non}
V.~H{\"a}berle, A.~Hauswirth, L.~Ortmann, S.~Bolognani, and F.~D{\"o}rfler,
  ``Non-convex feedback optimization with input and output constraints,''
  \emph{IEEE Control Systems Letters}, vol.~5, no.~1, pp. 343--348, 2020.

\bibitem{nocedal1999}
J.~Nocedal and S.~J. Wright, \emph{\BIBforeignlanguage{eng}{Numerical
  optimization}}, ser. Springer series in operations research.\hskip 1em plus
  0.5em minus 0.4em\relax New York: Springer, 1999.

\bibitem{colombino2019towards}
M.~Colombino, J.~W. Simpson-Porco, and A.~Bernstein, ``Towards robustness
  guarantees for feedback-based optimization,'' in \emph{58th Conference on
  Decision and Control (CDC)}, 2019, pp. 6207--6214.

\bibitem{chan2025regularization}
W.~Chan, Z.~He, K.~Moffat, S.~Bolognani, M.~Muehlebach, and F.~Dörfler,
  ``Robust feedback optimization with model uncertainty: A regularization
  approach,'' \emph{arXiv:2503.24151 [math.OC]}, 2025.

\bibitem{he2022model}
Z.~He, S.~Bolognani, J.~He, F.~D{\"o}rfler, and X.~Guan, ``Model-free nonlinear
  feedback optimization,'' \emph{IEEE Transactions on Automatic Control},
  vol.~69, no.~7, pp. 4554--4569, 2024.

\bibitem{lofberg2004yalmip}
J.~Lofberg, ``Yalmip: A toolbox for modeling and optimization in matlab,'' in
  \emph{IEEE International Conference on Robotics and Automation}, 2004, pp.
  284--289.

\bibitem{belgioioso2023multiarea}
G.~Belgioioso, S.~Bolognani, G.~Pejrani, and F.~D{\"o}rfler, ``Tutorial on
  congestion control in multi-area transmission grids via online feedback
  equilibrium seeking,'' in \emph{62nd IEEE Conference on Decision and Control
  (CDC)}, 2023.

\end{thebibliography}


\end{document}